\documentclass[twocolumn,aps,prl,superscriptaddress,showpacs]{revtex4}
\usepackage{amssymb}
\usepackage{graphicx}
\usepackage{amsmath}
\usepackage{amsfonts}
\usepackage{color}
\usepackage{ulem}

\usepackage{hyperref}

\begin{document}
\title{Symmetry-protected non-Abelian braiding of Majorana Kramers' pairs}
\author{Pin Gao}
\affiliation{International Center for Quantum Materials and School of Physics, Peking University, Beijing 100871, China}
\affiliation{Collaborative Innovation Center of Quantum Matter, Beijing 100871, China}
\author{Ying-Ping He}
\affiliation{International Center for Quantum Materials and School of Physics, Peking University, Beijing 100871, China}
\affiliation{Collaborative Innovation Center of Quantum Matter, Beijing 100871, China}
\author{Xiong-Jun Liu\footnote{Correspondence should be addressed to: xiongjunliu@pku.edu.cn}}
\affiliation{International Center for Quantum Materials and School of Physics, Peking University, Beijing 100871, China}
\affiliation{Collaborative Innovation Center of Quantum Matter, Beijing 100871, China}
\date{\today}

\begin{abstract}
We develop the complete theory for non-Abelian braiding of Majorana Kramers' pairs (MKPs) in time-reversal (TR) invariant topological superconductors. By introducing an effective Hamiltonian approach to describe the braiding of MKPs, we show that the non-Abelian braiding is protected when the effective Hamiltonian exhibits a new TR like anti-unitary symmetry, which is satisfied if the system is free of dynamical noise. Importantly, even the dynamical noise may not cause error in braiding, unless the noise correlation function breaks a dynamical TR symmetry, which generalizes the TR symmetry protection of MKPs to dynamical regime. Moreover, the resulted error by noise is shown to be a higher order effect, compared with the decoherence of Majorana qubits without TR symmetry protection. These results show that the non-Abelian braiding of MKPs is observable and may have versatile applications to future quantum computation technologies.
\end{abstract}

\maketitle

{\it Introduction}.--Majorana zero modes (MZMs) are self-hermitian quasiparticles which can exist in the ends of a one-dimensional (1D) topological superconductor (TSC)~\cite{Kitaev1} and the vortex cores of a 2D TSC~\cite{Read}. Due to the self-hermitian property, a single MZM has no well defined Hilbert space spanned by usual complex fermion quantum states. Instead, a complex fermion mode, whose Hilbert space defines a single qubit and is spanned by two fermionic quantum states $|0\rangle$ and $|1\rangle$, can be formed by two independent Majorana quasiparticles. This follows that a single Majorana mode has an irrational quantum dimension $\sqrt{2}$~\cite{Nayak}. The non-integer quantum dimension leads to an exotic property, namely, the non-Abelian statistics for the MZMs~\cite{Nayak1996,Ivanov,Sankar,Jason2011}, which is the essential motivation in the recent years of extensive studies of TSCs in condensed matter physics, see e.g. refs.~\cite{Fu1,Sau0,Alicea,Roman0,Oreg,Kouwenhoven,Deng,Das,Yazdani,Liu,Patrick2012PRL,Xin2016}.

The non-Abelian statistics of MZMs has been mostly considered in the chiral (class D) TSCs, where the isolated MZMs can be braided to demonstrate non-Abelian operations. In contrast, in a time-reversal (TR) invariant (class DIII) TSC, the MZMs come in pairs, referred to as Majorana Kramers' pairs (MKPs) due to the Kramers¡¯ theorem~\cite{Qi,Teo,Schnyder,Beenakker,Law2,Nagaosa,Kane,Keselman2013}. The TR symmetry protection is an essential ingredient in the DIII class TSCs, which generates many interesting new physics. Especially, it was proposed recently that the braiding of MKPs is non-Abelian~\cite{LiuPRX2014}, and thus may be applied to quantum computation~\cite{Oreg2014,Gaidamauskas2014,FanZhang2014,Loss1,Loss2,Sato2014,Flensberg2016,Flensberg2014}. Nevertheless, while the MKPs are protected by TR symmetry, by definition the non-Abelian braiding of them excludes the local operations on each single MKP~\cite{LiuPRX2014}, which cannot be achieved solely by the TR symmetry protection~\cite{Flensberg2014}. Thus the important fundamental and realistic questions arise. First of all, what are the sufficient symmetry conditions for the non-Abelian braiding of MKPs? Moreover, how well such conditions can be satisfied in the proposed typical TR invariant TSCs? Answering the two questions shall complete the definition of symmetry protected non-Abelian braiding of MKPs and show the potential feasibility of applying such new type of non-Abelian physics to future computation technologies.

In this work, we develop the complete theory for non-Abelian statistics of MKPs in TR invariant TSCs. We find that the ideal non-Abelian braiding of MKPs is protected when the effective Hamiltonian, introduced to describe MKPs' braiding, exhibits a new TR like anti-unitary symmetry. We show that this symmetry can be well satisfied except that the system has dynamical noise which breaks the dynamical TR symmetry, but results in only a higher-order error in the braiding, compared with the decoherence of Majorana qubits in chiral TSCs caused by dynamical perturbations. We confirm our findings with analytical and numerical results.

{\it MKPs' braiding}.--A MKP can exist at an interface between topological and trivial regions of a 1D TR invariant TSC. Shifting the topological and trivial interfaces through tuning chemical potential can transport MKPs, as an essential step to braid MKPs~\cite{Jason2011,LiuPRX2014}. While the non-Abelian statistics of MKPs are fundamental and model independent, without loss of generality, for the concrete study we consider a generic single-band 1D TR invariant TSC described by
\begin{eqnarray}\label{braiding}
H_0&=&\sum_{\left<i,j\right>,\sigma}\! t_0 c_{i\sigma}^\dagger c_{j\sigma}+\sum_{j}\!(\pm\alpha_R c_{j\uparrow}^\dagger c_{j\pm1\downarrow}\!+\Delta_{p} c_{j\uparrow}c _{j+1\uparrow}\nonumber\\
&&+\Delta^*_{p} c_{j\downarrow}c _{j+1\downarrow}+\Delta_s c_{j\uparrow} c_{j\downarrow}\!+\!h.c.)\!-\!\mu\!\sum_{j \sigma} \!n_{j \sigma},
\end{eqnarray}
where $\sigma=\uparrow, \downarrow$, $\mu$ is chemical potential, and $t_0$ denotes nearest-neighbor hopping between $i$ and $j$ sites. The $p$-wave ($\Delta_p$) and $s$-wave ($\Delta_s$) pairing orders, and Rashba SO coupling ($\alpha_R$) have been taken into account. The system is in topological regime for $|\Delta_p|>|\Delta_s|$ and $|\mu|<2t_0$, while it is trivial if $|\mu|>2t_0$~\cite{LiuPRX2014}. The total braiding Hamiltonian can be written as $H(t)=H_0+\delta H(t)$, with the time-dependent term $\delta H(t)$ denoting the braiding manipulation which can be achieved by tuning chemical potential. By definition the non-Abelian statistics of MKPs exclude local operations on each MKP~\cite{LiuPRX2014}, and necessitate symmetry protection. Below we show first the sufficient conditions for non-Abelian statistics of MKPs.

The first condition is the TR symmetry. Note that the braiding of MKPs can be physically performed by tuning gate in a trijunction formed by 1D TR invariant TSCs. The process can be described by the total braiding Hamiltonian $H(t)$, where the time $-T/2\leq t\leq T/2$ with $T$ the braiding period. To ensure that the system has MKPs during the braiding process, we require that the Hamiltonian satisfies TR symmetry at each time, namely,
\begin{eqnarray}\label{TRsymmetry}
{\cal T}H(t){\cal T}^{-1}=H(t),
\end{eqnarray}
where ${\cal T}=i\sigma_y{\cal K}$, with ${\cal K}$ being the complex conjugate and Pauli matrix $\sigma_y$ acting on spin. Under this condition the instantaneous MKPs exist in the ends or the interface between topological and trivial regions of the TR invariant TSC. The two MKPs under braiding are denoted as $\gamma_{1}(t),\tilde\gamma_1(t)$ and $\gamma_2(t),\tilde\gamma_2(t)$.

The second condition for the non-Abelian braiding is the adiabatic condition, which is necessary for generic non-Abelian braiding of MZMs rather than solely for the MKPs~\cite{Jason2011}, and is quantified by $T\gg1/E_g$, with $E_g$ the bulk gap of the TSC. Under this condition, the Majorana modes are decoupled from the excitations beyond the bulk gap. Then the braiding operation can be captured by the evolution in the Majorana subspace, given by
%\begin{eqnarray}\label{evolution}
$\{|\gamma_j(t)\rangle,|\tilde\gamma_j(t)\rangle\}^T=\hat Te^{-i\int_{-T/2}^t H(\tau)d\tau}\{|\gamma_j(0)\rangle,|\tilde\gamma_j(0)\rangle\}^T.$
%\end{eqnarray}
Here $\hat T$ represents the time-ordered integral. In the adiabatic regime, we can facilitate our study by introducing an effective Hamiltonian to describe the braiding process
\begin{eqnarray}\label{effective}
H_E(T)&=&iT^{-1}\log\bigr[\hat Te^{-i\int_{-T/2}^{T/2}d\tau H(\tau)}\bigr]\nonumber\\
&=&iT^{-1}\lim_{\Delta t\to 0}\log\bigr[e^{-iH(T/2)\Delta t }e^{-i H(T/2-\Delta t)\Delta t}\nonumber\\
&&\cdots e^{-i H(-T/2+\Delta t)\Delta t} e^{-iH(-T/2)\Delta t }\bigr].
\end{eqnarray}
Then the braiding operation reads
$U_{12}(T)=e^{-iH_E(T)T}$.
To discover the sufficient symmetry condition of the original Hamiltonian $H(t)$, a difficult task, for the non-Abelian braiding can be reduced to study the symmetry condition of the effective one $H_E(T)$, which can be represented by the four Majorana modes. In the following we shall see the explicit advantages of the effective Hamiltonian approach proposed here.

We proceed to introduce the last symmetry condition for the non-Abelian statistics. From the relation~\eqref{effective} we note that in general the symmetries respected by $H$ and $H_E$ are different. It can be shown that the charge conjugation symmetry, sending the Hamiltonian to ${\cal C}H(t){\cal C}^{-1}=-H^*(t)$, is also respected by $H_E$ as ${\cal C}H_E(T){\cal C}^{-1}=-H_E^*(T)$. However, the TR symmetry in Eq.~\eqref{TRsymmetry} is generically broken in $H_E$ according to Eq.~\eqref{effective}~\cite{SI}. Our key purpose is to introduce a new TR like symmetry which is respected by the effective Hamiltonian. If we define a Majorana swapping operator $\hat S$ as $\hat S\gamma_{1(2)}\hat S^{-1}=\gamma_{2(1)}$ and $\hat S\tilde\gamma_{1(2)}\hat S^{-1}=\tilde\gamma_{2(1)}$, and it sends that $\hat SH(-t)\hat S^{-1}=H(t)$, we can easily prove that $H_E(T)$ satisfies
\begin{eqnarray}\label{swappingsymmetry}
\Theta H_E(T)\Theta^{-1}=H_E(T),
\end{eqnarray}
where $\Theta={\cal T}\hat S$ is a TR like anti-unitary operator. With this symmetry condition the non-Abelian braiding of MKPs can be rigorously obtained.

Together with the charge conjugation and the new TR like symmetries, the effective Hamiltonian must take the following generic form
\begin{eqnarray}
H_E=i\epsilon_1\gamma_1\tilde{\gamma}_1-i\epsilon_1\gamma_2\tilde{\gamma}_2+i\epsilon_2\gamma_1\gamma_2+i\epsilon_2\tilde{\gamma}_1\tilde{\gamma}_2,
\end{eqnarray}
where $\epsilon_{1,2}$ are parameters to be determined by the definition of braiding. Then, in the basis $\{\gamma_1,\tilde\gamma_1,\gamma_2,\tilde\gamma_2\}$ the operation for the single braiding is obtained by
$$U_{12}=\left(\begin{matrix}
\cos \epsilon  T&\frac{\epsilon_1\sin \epsilon  T}{\epsilon }&\frac{\epsilon_2\sin \epsilon  T}{\epsilon }&0\\
-\frac{\epsilon_1\sin \epsilon  T}{\epsilon }&\cos \epsilon  T&0&\frac{\epsilon_2\sin \epsilon  T}{\epsilon }\\
-\frac{\epsilon_2\sin \epsilon  T}{\epsilon }&0&\cos \epsilon  T&-\frac{\epsilon_1\sin \epsilon  T}{\epsilon }\\
0&-\frac{\epsilon_2\sin \epsilon  T}{\epsilon }&\frac{\epsilon_1\sin \epsilon  T}{\epsilon }&\cos \epsilon  T
\end{matrix}\right),$$
with $\epsilon=(\epsilon_1^2+\epsilon_2^2)^{1/2}$.
Note that braiding exchanges MKP positions $\gamma_1,\tilde{\gamma}_1\leftrightarrow\gamma_2,\tilde{\gamma}_2$, which forces $\cos \epsilon  T=0$ and $\epsilon_1=0$. It follows immediately that $\sin\epsilon T=1$ (or $-1$) and $U_{12}\gamma_j=\mathcal{B}_{12}\gamma_j\mathcal{B}_{12}^{-1}$ (similar for $\tilde\gamma_j$), where
\begin{equation}\label{braidingoperator}
\mathcal{B}_{12}=\exp(-\frac{\pi}{4}\gamma_1\gamma_2)\exp(-\frac{\pi}{4}\tilde{\gamma}_1\tilde{\gamma}_2)
\end{equation}
renders the symmetry-protected non-Abelian braiding of MKPs~\cite{LiuPRX2014}. It can be seen that the above proof is not restricted by any specific Hamiltonian. Thus the non-Abelian statistics of MKPs are valid if only the above symmetry conditions are satisfied.
\begin{figure}[t]
\includegraphics[width=0.9\columnwidth]{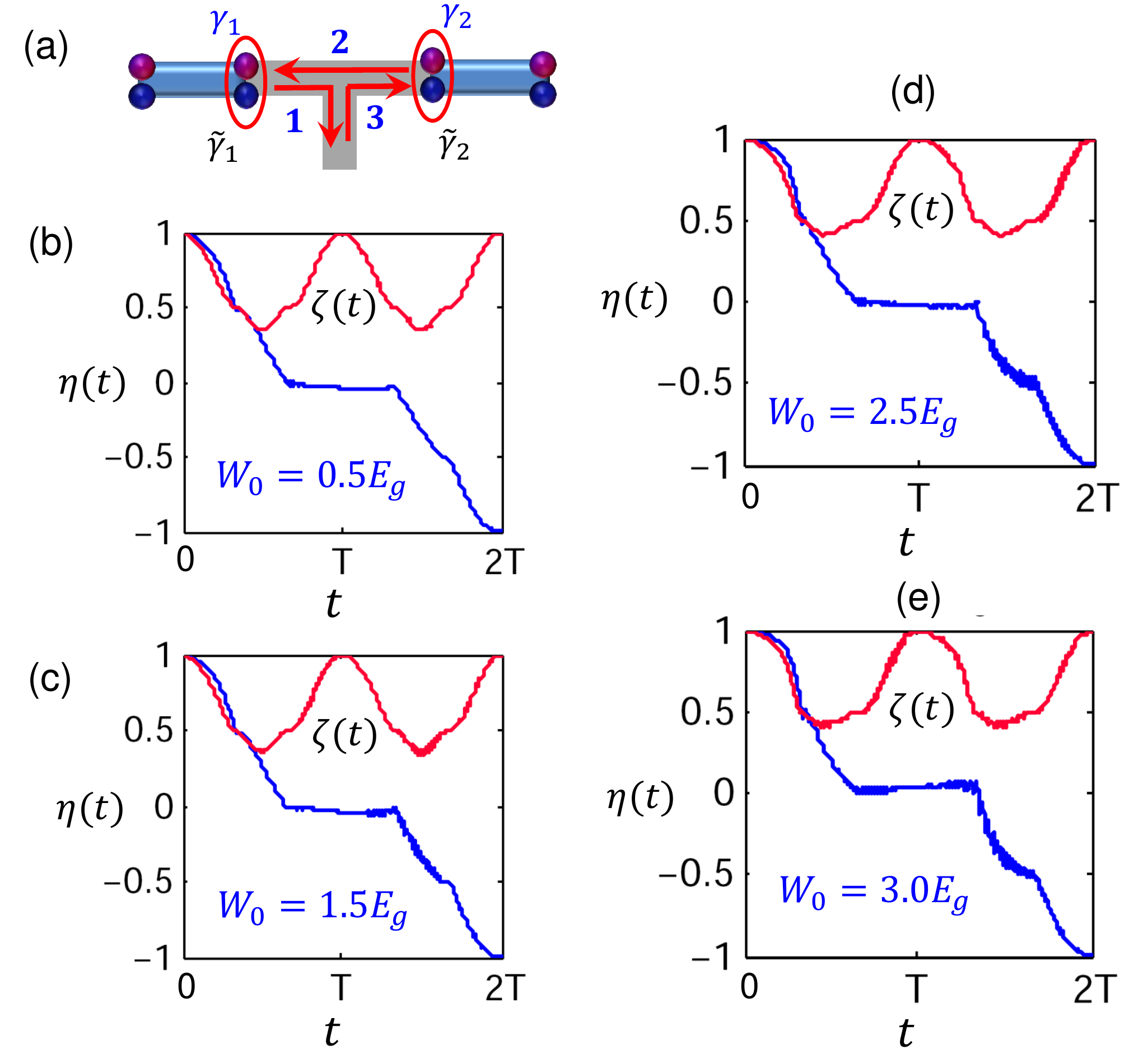} \caption{(Color online) Non-Abelian braiding of MKPs immune to static disorder scattering. (a) Sketch of the braiding process, with $1,2,3$ denoting the sequence of transporting MKPs by gate control. (b-d) Evolution of MKP wave function in a full braiding with different disorder strengths $W_0$ (in unit of bulk gap $E_g$). The non-Abelian braiding of MKPs is confirmed by that the magnitude $\eta(t)=\langle\gamma_1(0)|\gamma_1(t)\rangle|_{t=2T}=-1$ after a full braiding at $t=2T$ (similar for other Majorana modes)~\cite{Jason2011,Sato2015,Dongling}. The adiabatic condition is satisfied in that $\zeta(t)=\sum_{j=1,2}[|\langle\gamma_1(t)|\gamma_j(0)\rangle|^2+|\langle\gamma_1(t)|\tilde\gamma_j(0)\rangle|^2]$ returns to unity after a single ($t=T$) and full ($t=2T$) braiding.}
\label{disorder}
\end{figure}

{\it Static disorder effect.}--After presenting the generic symmetry protection of non-Abelian statistics of MKPs, it is instructive to know how well these conditions be satisfied in the typical TR invariant TSCs. We consider the following situation that, the time-dependence of the braiding Hamiltonian $\delta H(t)$ solely comes from the adiabatic parameter manipulation, e.g. the gate control, for the braiding. The system can have static disorders, but has no dynamical noise. In this case, during the braiding the statistic disorder can in general bring about local rotation on the MKPs, with, however, the rotating angles for $\gamma_1,\tilde\gamma_1$ and $\gamma_2,\tilde\gamma_2$ being opposite (details are found in the supplementary material~\cite{SI}). Thus, a generic Majorana swapping symmetry exists, transforming the Hamiltonian by $SH(-t)S^{-1}=H(t)$ and MKPs according to $\hat S\gamma_{1/2}\hat S^{-1}=\cos(\pm\theta_b)\gamma_{2/1}+\sin(\pm\theta_b)\tilde\gamma_{2/1}$, and $\hat S\tilde\gamma_{1/2}\hat S^{-1}=\cos(\pm\theta_b)\tilde\gamma_{2/1}-\sin(\pm\theta_b)\gamma_{2/1}$. Here the rotating angle $\theta_b$ is system dependent. In particular, if the braiding Hamiltonian is inversion symmetric with respect to junction position, one has $\theta_b=0$. The braiding operator is then given by ${\cal B}_{12}=\exp(-\pi\gamma_1\gamma_2'/4)\exp(-\pi\tilde\gamma_1\tilde\gamma_2'/4)$ with $\gamma_2'=\cos\theta_b\gamma_2+\sin\theta_b\tilde\gamma_2$ and $\tilde\gamma_2'={\cal T}\gamma_2'{\cal T}^{-1}$, recovering the non-Abelian braiding.

The above results show a remarkable feature: while the non-Abelian braiding of MKPs necessitates symmetry protection, it can be physically well satisfied if the 1D TR invariant TSC has no dynamical noise. These predictions are confirmed by the numerical simulations shown in Fig.~\ref{disorder}, based on the 1D TSC described by Eq.~\eqref{braiding} and taking into account the random static disorder potential \begin{eqnarray}\label{disorder}
V_{\rm dis}&=&\sum_{j}W_j(n_{j \uparrow}+n_{j\downarrow}),
\end{eqnarray}
with disorder strength $W_j$ randomly distributed within the range of $[-W_0,W_0]$.
The hopping, Rashba, $p$-wave and $s$-wave pairing strengths are rescaled to be dimensionless and taken as $t_0=10, \Delta_s=1$, $\Delta_p=2$, $\alpha_R=1$. In the topological region the chemical potential is set as $\mu=7$, with the topological gap without disorder being $E_g\approx 0.58$, and the braiding through the junction is further
performed by locally tuning $\mu$ [Fig.~\ref{disorder}(a)]. The non-Abelian braiding is precisely confirmed for all disorder strengths without destroying the bulk topology~\cite{SI}. Increasing $W_0$ induces more fluctuations in the Majorana wave functions in the intermediate evolution, but does not affect the result after braiding [Fig.~\ref{disorder}(b-e)].

{\it Dynamical noise}.--Now we proceed to study the effect of dynamical noise on the braiding. The dynamical noise may induce random local operations on a single MKP, leading to the error of the non-Abelian braiding, if the dynamical noise breaks certain symmetries. For a generic study, we consider a MKP in a 1D TR invariant TSC, denoted as $\gamma_a$ and $\tilde\gamma_a$, coupled to the bulk fermionic modes, denoted as $c_j$ and with energy $\epsilon_{j}$, by a dynamical perturbation. The Hamiltonian can be generically described by $H=H_0+H_{p}$, with
\begin{eqnarray}
H_{0}&=&\sum_{j}\epsilon_{j}(c_{j}^{\dagger}c_{j}+\tilde{c}_{j}^{\dagger}\tilde{c}_{j})\nonumber\\
H_{p}&=&\gamma_a \sum_{j} (V_{j1} c_{j}-V_{j1}^{*} c_{j}^{\dagger} +V_{j2} \tilde{c}_{j}-V_{j2}^{*} \tilde{c}_{j}^{\dagger}) + {\rm T.P.}\nonumber
\end{eqnarray}
where the fermionic modes $\tilde{c}_{j}=\mathcal{T}c_{j}\mathcal{T}^{{-1}}$ and ${\rm T.P.}$ means TR part of the former term in $H_{p}$. The Hamiltonian $H_{p}$ describes the couplings between bulk fermionic modes and the MKP by the dynamical noise $V_{jm}(t)$. The correlation function of the dynamical noise satisfies
$\left\langle V_{j1}(t_1)V_{j2}(t_2) \right\rangle_0=V_0^2\mathcal{C}_{j}(t_1-t_2)$,
where $V_0^2$ characterizes the noise strength and $\mathcal{C}_{j}(\tau)$ determines the property of the correlation function~\cite{SI,Zoller2004}. Besides, we take the configuration averaging to be $\left\langle V_{jm} \right\rangle_0=0$, since any nonzero value of this term can be absorbed by redefining the energy of fermionic states.

We treat the dynamical noise beyond perturbation. By including all order contributions, the amplitude of transition from one Majorana ($\gamma_a$) to another ($\tilde\gamma_a$), calculated by $\chi(t)=\langle\tilde{\gamma}_a |\hat T\exp[-i\int_{-T/2}^tH(\tau)d\tau]|\gamma_a\rangle$, is given by
\begin{eqnarray}\label{ftran}
\chi(t)&=&2V_0^2\sum_j \int_{-T/2}^{t}d \tau_{1}\int_{-T/2}^{\tau_1}d \tau_{2}\ \Re\bigr\{[\mathcal{C}_j(\tau_1-\tau_2)\nonumber\\
&&-\mathcal{C}_j(\tau_2-\tau_1)]e^{i\epsilon_j(\tau_1-\tau_2)}\bigr\}+\chi^{(4)}(t)+\cdots,
\end{eqnarray}
where $\chi^{(j)}$, with $j$ being even, denotes the $j$-th order contribution~\cite{SI}.
The above formula exhibits two important features for the dynamical noise. First, if the noise correlation respects a dynamical TR symmetry in the time domain,
$\mathcal{C}_j(\tau)=\mathcal{C}_j(-\tau)$, the coupling $\chi(t)$ between $\gamma_a$ and $\tilde\gamma_a$ vanishes for all order contributions~\cite{SI}, and the Majorana swapping symmetry is
guaranteed. This is a profound result, showing that {\it the mixing in a MKP must break either the static TR symmetry (${\cal T}$) or the dynamical TR symmetry defined via the correlation function of dynamical noise, and generalizing the TR symmetry protection of MKPs to the dynamical regime.}
Moreover, the leading transition probability in a MKP is proportional to $|\chi(t)|^2\propto V_0^4$, which implies that the leading order contribution due to dynamical noise is a {\it second-order} transition, describing the process that a Majorana mode ($\gamma_a$) virtually transitions to a bulk fermionic state, and then back to its TR partner ($\tilde\gamma_a$). This result is deeply related to the TR symmetry of the original DIII class TSC, which excludes the first-order direct coupling in a MKP. Note that the dynamical noise can also bring about decoherence effect in the Majorana qubits in a D-class chiral TSC without TR symmetry, where the leading contribution comes from the first-order transition~\cite{Goldstein2011}. The high-order contribution in MKPs implies that while the dynamical noise is detrimental, the caused error can be generically reduced if the noise strength is not strong.

For a quantitative study, we consider the Hamiltonian~\eqref{braiding} for the TSC and a dynamical noise in the form
\begin{eqnarray}
%\begin{split}
H_{p}=\sum_{j}V_j[\cos(\omega t)(c_{j\uparrow} c_{j\downarrow}\!+\!h.c.)\!-\!\cos (\omega t+\delta \phi_j)n_j],
%\end{split}
\end{eqnarray}
which describes dynamical fluctuations of frequency $\omega$ in the $s$-wave pairing and chemical potential (other parameter fluctuations are similar), with $n_j=n_{j\uparrow}+n_{j\downarrow}$. The phase difference $\delta \phi_j$ in fluctuating chemical potential and $s$-wave pairing is assumed to be position dependent. For simplicity we take the amplitude $V_j\equiv V_0$, while in reality a random distribution in position is allowed and can further reduce the error caused by dynamical noise~\cite{SI}, like the effect of random phase distribution $\delta \phi_j$ to be discussed below. It is easy to see that if $\delta\phi_j=0$ over the system, one has $H_p(t)=H_p(-t)$. Then according to previous generic discussion the coupling in a MKP vanishes and the local rotations disappear.

\begin{figure}[h]
\includegraphics[width=0.9\columnwidth]{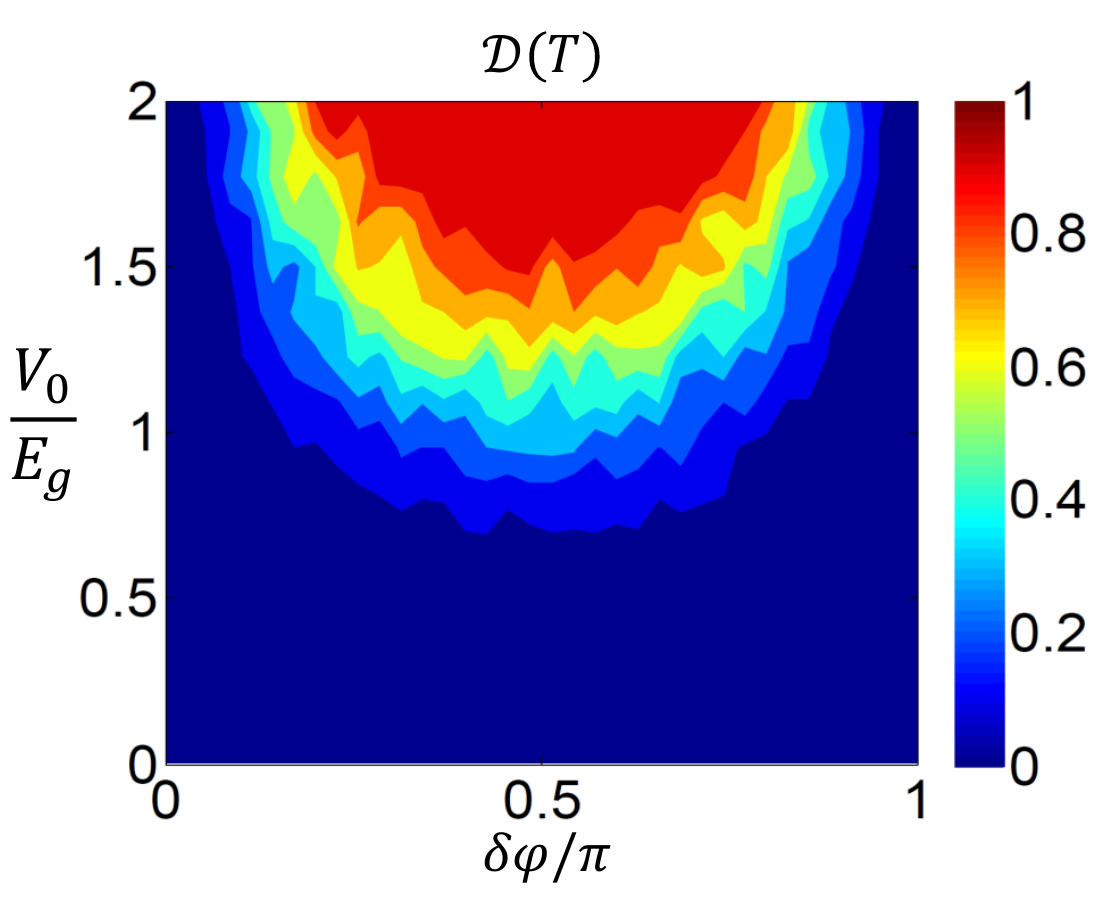} \caption{(Color online) Numerical simulation of error in braiding $\mathcal{D}(T)$ induced by uniform dynamical noise with strength $V_0$. The bulk gap $E_g=0.58$. Note that for $\delta\varphi=0$ and $\pi$, the local rotations and braiding error disappear.}\label{phi}
\label{noise1}
\end{figure}
\begin{figure}[h]
\includegraphics[width=1.0\columnwidth]{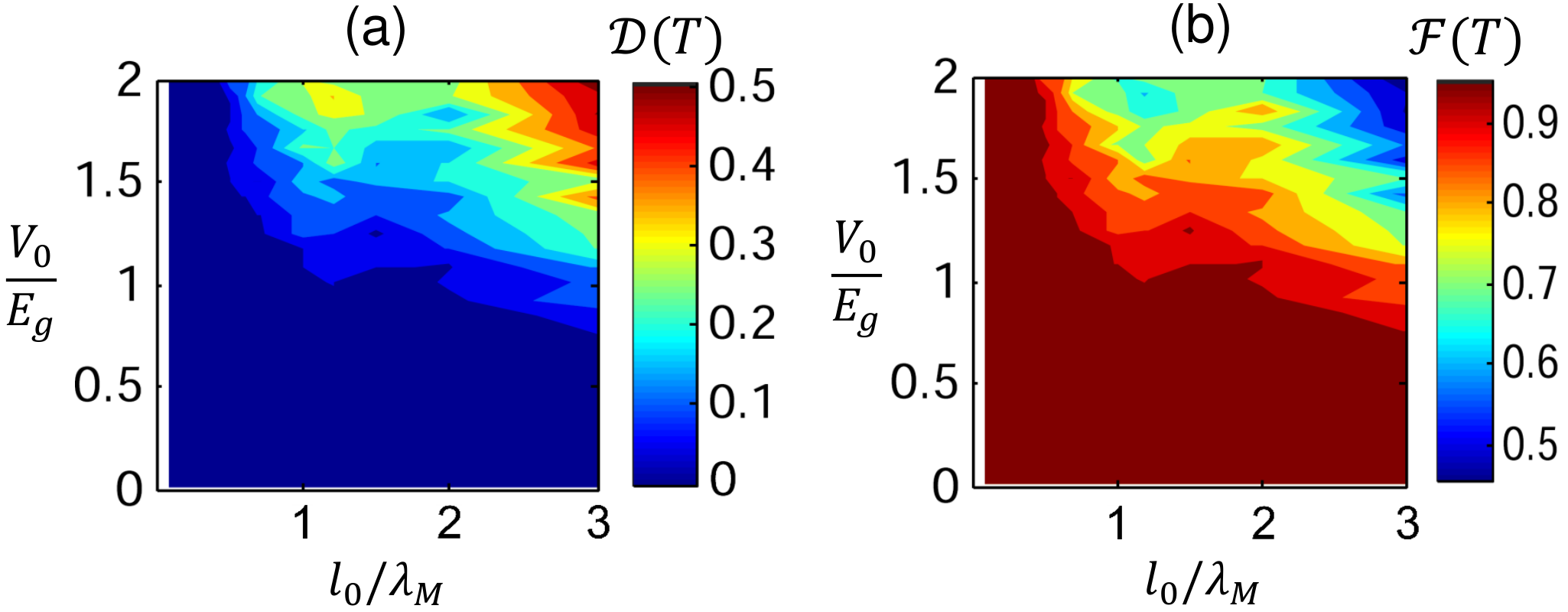} \caption{(Color online) (a) Error in braiding $\mathcal{D}(T)$ induced by random dynamical noise, versus disorder strength $V_0$ and coherence length $l_0$. (b) The fidelity of the braiding
${\cal F}(T)=|\langle\gamma_a|e^{-iH_ET}|\gamma_a\rangle|^2$. The gap $E_g\approx0.58$ and Majorana localization length $\lambda_M\sim20$ sites in the numerical simulation.}
\label{noise2}
\end{figure}

The numerical results are shown for uniform $\delta\phi_j\equiv\delta\varphi$ (Fig.~\ref{noise1}) and random (Fig.~\ref{noise2}) phase distributions. The parameters of $H_0$ are the same as those used for numerical simulation in Fig.~\ref{disorder}, with the
localization length of MKP $\lambda_M\approx 20$ sites~\cite{note}. The noise frequency is set as $\omega=3/T\ll E_g$, with $T=100$. The error or local rotation can be quantified by $\mathcal{D}\!=\!|\left\langle\tilde{\gamma}_a|e^{-iH_ET}|\gamma_a\right\rangle|^2$. It can be read that $\mathcal{D}\to 0$ as $\delta \phi\to 0,\pi$ for uniform noise (Fig.~\ref{noise1}), consistent with our above analysis. In the weak noise regime we find that the results can be approximated by $\mathcal{D}(T)\approx 1-\cos^2\bigr(\frac{V_0^2}{4E_g^2}\langle\gamma_a|\sin\delta\phi_j|\gamma_a\rangle)$, giving
\begin{eqnarray}
\mathcal{D}\approx\frac{V_0^4}{16E_g^4}\langle\gamma_a|\sin\delta\phi_j|\gamma_a\rangle^2
-\frac{V_0^8}{763E_g^8}\langle\gamma_a|\sin\delta\phi_j|\gamma_a\rangle^4.
\end{eqnarray}
Here $\langle\gamma_a|\sin\delta\phi_j|\gamma_a\rangle$ denotes the average of random noise phase $\delta\phi_j$ experienced by the Majorana mode
$\gamma_a$. The randomness of the noise can be quantified by its spatial coherence length $l_0$, over which the phases at two positions are uncorrelated: $\langle\delta\phi_j\delta\phi_{j'}\rangle_{|j-j'|>l_0}=0$. As shown in Fig.~\ref{noise2}(a,b), the effect of dynamical noise is greatly reduced (or the fidelity of braiding is largely enhanced) when the noise coherence length $l_0$ is less than the MKP localization length $\lambda_M$. This confirms an important result: the non-Abelian braiding of MKPs is restored if the the random dynamical noise is spatially averaged to be zero in the Majorana localization length.

{\it Conclusions}.--In conclusion, we have developed a complete theory for symmetry protected non-Abelian statistics of MKPs in TR invariant TSCs. While necessitating symmetry protection, the non-Abelian braiding of MKPs is shown to be stable against static disorder scatterings, and be protected even in the presence of dynamical noise, given that the noise correlation function keeps a dynamical TR symmetry. The results reveal a novel generalization of the TR symmetry protection of MKPs to dynamical regime. Moreover, we have shown that in general the dynamical noise can at most bring about a higher order error to the braiding, which is deeply related to the symmetry protection of MKPs in the TR invariant TSC. Besides the manipulation of MKPs via braiding, the stability of the MKPs against disorder scatterings implies that one can also manipulate a single MKP by tuning local parameters in a controllable fashion. This enables rich operations of Majorana qubits formed by MKPs, and may open versatile applications to future quantum computation technologies, in particular, toward the realization of universal quantum computations.

We thank T. K. Ng, K. T. Law, and B. Liu for helpful discussions. This work
is supported by MOST (Grant No. 2016YFA0301600), NSFC (No. 11574008), and Thousand-Young-Talent Program of China.

%%%%%%%%%%%%%%%%%%%%%%%%%%%%%%%%%%%%%%%%%%%%%

\noindent

\onecolumngrid

%%%%%%%%%%%%%%%%%%%%%%%%%%%%%%%%%%%%%%
%%   Supplementary Information
%%%%%%%%%%%%%%%%%%%%%%%%%%%%%%%%%%%%%%
%\appendix
\renewcommand{\thesection}{A-\arabic{section}}
\setcounter{section}{0}  %  this will re-count section from 1
\renewcommand{\theequation}{A\arabic{equation}}
\setcounter{equation}{0}  %  this will re-count eq from 1
\renewcommand{\thefigure}{A\arabic{figure}}
\setcounter{figure}{0}  %  this will re-count eq from 1

\indent

\section*{\large Supplementary Material}

\section*{I. Effective Hamiltonian approach}

\subsection*{A. Definition}
The effective Hamiltonian is defined through the the evolution operator by $\exp[-iH_ET]\equiv T^{-1}e^{-i\int_{-T/2}^{T/2}H(\tau)d\tau}$, giving that
\begin{eqnarray}
H_E(T)&=&iT^{-1}\log\bigr[\hat Te^{-i\int_{-T/2}^{T/2}d\tau H(\tau)}\bigr]\nonumber\\
&=&iT^{-1}\lim_{\Delta t\to 0}\log\bigr[e^{-iH(T/2)\Delta t }e^{-i H(T/2-\Delta t)\Delta t}\nonumber\\
&&\cdots e^{-i H(-T/2+\Delta t)\Delta t} e^{-iH(-T/2)\Delta t }\bigr],
\end{eqnarray}
where $\Delta t=T/N\rightarrow0$ with $N\rightarrow\infty$. In the following we investigate the symmetry properties of the effective Hamiltonian and study the sufficient conditions for the non-Abelian braiding of Majorana Kramers' pairs (MKPs).

Note that we consider a time-dependent Hamiltonian. If at any time $t$, the Hamiltonian satisfies a symmetry as
$\hat{R}iH(t)\hat{R}^{-1}=siH(t)$,
where $\hat{R}$ does not depend on time, and $s=1$ or $-1$. If $H_E$ also satisfies the same symmetry
$$\hat{R}iH_E\hat{R}^{-1}=siH_E,$$
one can easily obtain that
\begin{eqnarray}
\hat{R}e^{-iH_ET}\hat{R}^{-1}&=&e^{-\hat{R}iH_E\hat{R}^{-1}T}\nonumber\\
&=&e^{-isH_ET}\nonumber\\
&=& \left\{ \begin{array}{ll}
e^{-iH_ET},&s=1,\\
(e^{-iH_ET})^{\dagger},&s=-1.
\end{array}\right.
\end{eqnarray}
Equivalently, the evolution operator $U(T)=\exp(-iH_ET)$ transforms as
\begin{eqnarray}\label{evolutionSI1}
U(T)= \left\{ \begin{array}{ll}
\hat{R}U(T)\hat{R}^{-1}, \ \ \ \ \  \mbox{for} \ \ s=1,\\
\bigr[\hat{R}U(T)\hat{R}^{-1}\bigr]^\dag,  \ \ \mbox{for} \ \ s=-1.
        \end{array} \right.
\end{eqnarray}

\subsection*{B. The case with $s=1$}

We consider first the case with $s=1$, in which case we shall see that any symmetry respected by $H(t)$ is also respected by the effective Hamiltonian $H_E$. Actually, we can check straightforwardly that
\begin{equation}
\hat{R}U(T)\hat{R}^{-1}=\lim_{N\to \infty}e^{-\hat{R}[iH(T/2-\Delta t)]\hat{R}^{-1}\Delta t }\cdots e^{-\hat{R}[i H(\Delta t-T/2)]\hat{R}^{-1}\Delta t} e^{-\hat{R}[iH(-T/2)]\hat{R}^{-1}\Delta t }=U(T).\label{s1}
\end{equation}
Then from Eq.~\eqref{evolutionSI1}, we know that the effective Hamiltonian also satisfies the same symmetry denoted by $\hat R$. In particular,
if $\hat{R}$ is an anti-unitary operator which has complex conjugate ${\cal K}$, one shall have $\hat{R}H(t)\hat{R}^{-1}=-H(t)$.
An example is that if the charge conjugation symmetry (denoted as ${\cal C}$) is satisfied by the original Hamiltonian $H(t)$ (note that in this case the definition of the symmetry should be slightly modified to be ${\cal C}H(t){\cal C}^{-1}=-H^*(t)$), it must be satisfied by the effective Hamiltonian.

\subsection*{C. The case with $s=-1$}

The situation with $s=-1$, which is the case for the time-reversal symmetry, is very different. It can be found that
\begin{equation}
\left[\hat{R}U(T)\hat{R}^{-1}\right]^\dagger=\lim_{N\to \infty}e^{-iH(-T/2)\Delta t }e^{-i H(\Delta t-T/2)\Delta t}\cdots e^{-iH(T/2-\Delta t)\Delta t },
\end{equation}
which can be different from the evolution operator $U(t)$. Thus the time-reversal symmetry, which is respected by $H(t)$, can generically be broken in the effective Hamiltonian $H_E$. To have a more concrete picture, we take an example by considering the following periodically driven Hamiltonian,
\begin{equation}
H(t)=H_0+2H_1\cos \omega t+2H_2\sin \omega t.
\end{equation}
We shall see that this Hamiltonian is relevant when the dynamical noise is taken into account. To show analytically how the symmetries are broken in the effective Hamiltonian, we consider the perturbation up to the order of $1/\omega^2$. The effective Hamiltonian can be obtained by Floquet theory
\begin{eqnarray}
H_E&=&H_0+\frac{1}{\omega}[V_1,V_{-1}]+\frac{1}{2\omega^2}\bigr\{[V_1,H_0],V_{-1}]+h.c.\bigr\}+\mathcal{O}(\frac{1}{\omega^3})\nonumber\\
&=&H_0+\frac{i}{\omega}[H_1,H_2]+\frac{1}{\omega^2}\bigr[2(H_1H_0H_1+H_2H_0H_2)\nonumber\\
&&+(H_1^2H_0+H_2^2H_0+h.c.)\bigr]+\mathcal{O}(\frac{1}{\omega^3})
\end{eqnarray}
where $V_1=H_1-iH_2$ and $V_{-1}=V_1^\dagger$. Note that we require $\hat{R}iH_{j}\hat{R}^{-1}=-iH_{j}$ for $s=-1$, with $j=0,1,2$, which follows that $\hat{R}iH(t)\hat{R}^{-1}+iH(t)=0$.
Thus we have
\begin{eqnarray}
\hat{R}iH_E\hat{R}^{-1}+iH_E=-\frac{2}{\omega}[H_1,H_2]+\mathcal{O}(\frac{1}{\omega^3}),
\end{eqnarray}
from which one finds immediately that if $[H_1,H_2]\neq 0$, the effective Hamiltonian $H_E$ breaks the symmetry $\hat R$, while it is satisfied in the original Hamiltonian $H(t)$.

\subsection*{D. Majorana swapping symmetry and non-Abelian braiding}

When the braiding Hamiltonian $H(t)$ respects a Majorana swapping symmetry, defined as $\hat S\gamma_{1(2)}\hat S^{-1}=\gamma_{2(1)}$ and $\hat S\tilde\gamma_{1(2)}\hat S^{-1}=\tilde\gamma_{2(1)}$, and the operator sends that $\hat SH(-t)\hat S^{-1}=H(t)$, we can find that
\begin{eqnarray}
\begin{array}{rcl}\vspace{3mm}
S\left[\hat{R}U(T)\hat{R}^{-1}\right]^\dagger S^{-1}&=&\lim_{N\to \infty}S e^{-iH(-T/2)\Delta t }e^{-i H(\Delta t-T/2)\Delta t}\cdots e^{-iH(T/2)\Delta t }S^{-1} \\\vspace{3mm}
&=&\lim_{N\to \infty}  e^{-iS H(-T/2)S^{-1}\Delta t }e^{-i S H(\Delta t-T/2)S^{-1}\Delta t}\cdots e^{-iS H(T/2)S^{-1}\Delta t } \\\vspace{3mm}
&=&\lim_{N\to \infty} e^{-iH(T/2)\Delta t }e^{-i H(T/2-\Delta t)\Delta t}\cdots e^{-iH(-T/2)\Delta t }\\\vspace{3mm}
&=&U(T).
\end{array}
\end{eqnarray}
Then the effective Hamiltonian satisfies a new TR like anti-unitary symmetry that
$\Theta iH_E \Theta^{-1}=-iH_E$, where $\Theta=\hat S\hat{R}$. Since $R$ is anti-unitary, we have $\Theta H_{E}\Theta ^{-1}=H_{E}$. For the present TR invariant topological superconductor, we take $\hat R$ to be the TR operator: $\hat{R}={\cal T}=i\sigma_y\mathcal{K}$.

In the presence of charge-conjugation and the new TR like symmetries, the effective Hamiltonian must take the following generic form
\begin{eqnarray}
H_E=i\epsilon_1\gamma_1\tilde{\gamma}_1-i\epsilon_1\gamma_2\tilde{\gamma}_2+i\epsilon_2\gamma_1\gamma_2+i\epsilon_2\tilde{\gamma}_1\tilde{\gamma}_2,
\end{eqnarray}
where $\epsilon_{1,2}$ are parameters to be determined by the definition of braiding. It can be seen that the effective Hamiltonian $H_E$ does not respect the TR symmetry ${\cal T}$, nor the $\hat S$ symmetry, but satisfies
\begin{eqnarray}
{\cal T}H_E{\cal T}^{-1}=-H_E, \ \ \hat SH_E\hat S^{-1}=-H_E, \ \ {\Theta}H_E\Theta^{-1}=H_E.
\end{eqnarray}
Consider the basis $\{\gamma_1,\tilde\gamma_1,\gamma_2,\tilde\gamma_2\}$, and the braiding matrix is given by $U_{12}=\exp(-iH_ET)$. By a straightforward calculation we obtain for the single braiding that
$$U_{12}=\left(\begin{matrix}
\cos \epsilon  T&\frac{\epsilon_1\sin \epsilon  T}{\epsilon }&\frac{\epsilon_2\sin \epsilon  T}{\epsilon }&0\\
-\frac{\epsilon_1\sin \epsilon  T}{\epsilon }&\cos \epsilon  T&0&\frac{\epsilon_2\sin \epsilon  T}{\epsilon }\\
-\frac{\epsilon_2\sin \epsilon  T}{\epsilon }&0&\cos \epsilon  T&-\frac{\epsilon_1\sin \epsilon  T}{\epsilon }\\
0&-\frac{\epsilon_2\sin \epsilon  T}{\epsilon }&\frac{\epsilon_1\sin \epsilon  T}{\epsilon }&\cos \epsilon  T
\end{matrix}\right).$$
The braiding exchange MKP positions $\gamma_1,\tilde{\gamma}_1\leftrightarrow\gamma_2,\tilde{\gamma}_2$, which forces $\cos \epsilon  T=0$ and $\epsilon_1=0$. It follows that
\begin{eqnarray}
U_{12}\gamma_j=\mathcal{B}_{12}\gamma_j\mathcal{B}_{12}^{-1}.
\end{eqnarray}
With this operation we find that a single braiding gives $\gamma_1\rightarrow\gamma_2,\tilde\gamma_1\rightarrow\tilde\gamma_2$, and $\gamma_2\rightarrow-\gamma_1,\tilde\gamma_2\rightarrow-\tilde\gamma_1$, rendering the symmetry-protected non-Abelian braiding of MKPs.

\subsection*{E. Static disorder}

\begin{figure}[h]
\includegraphics[width=0.6\columnwidth]{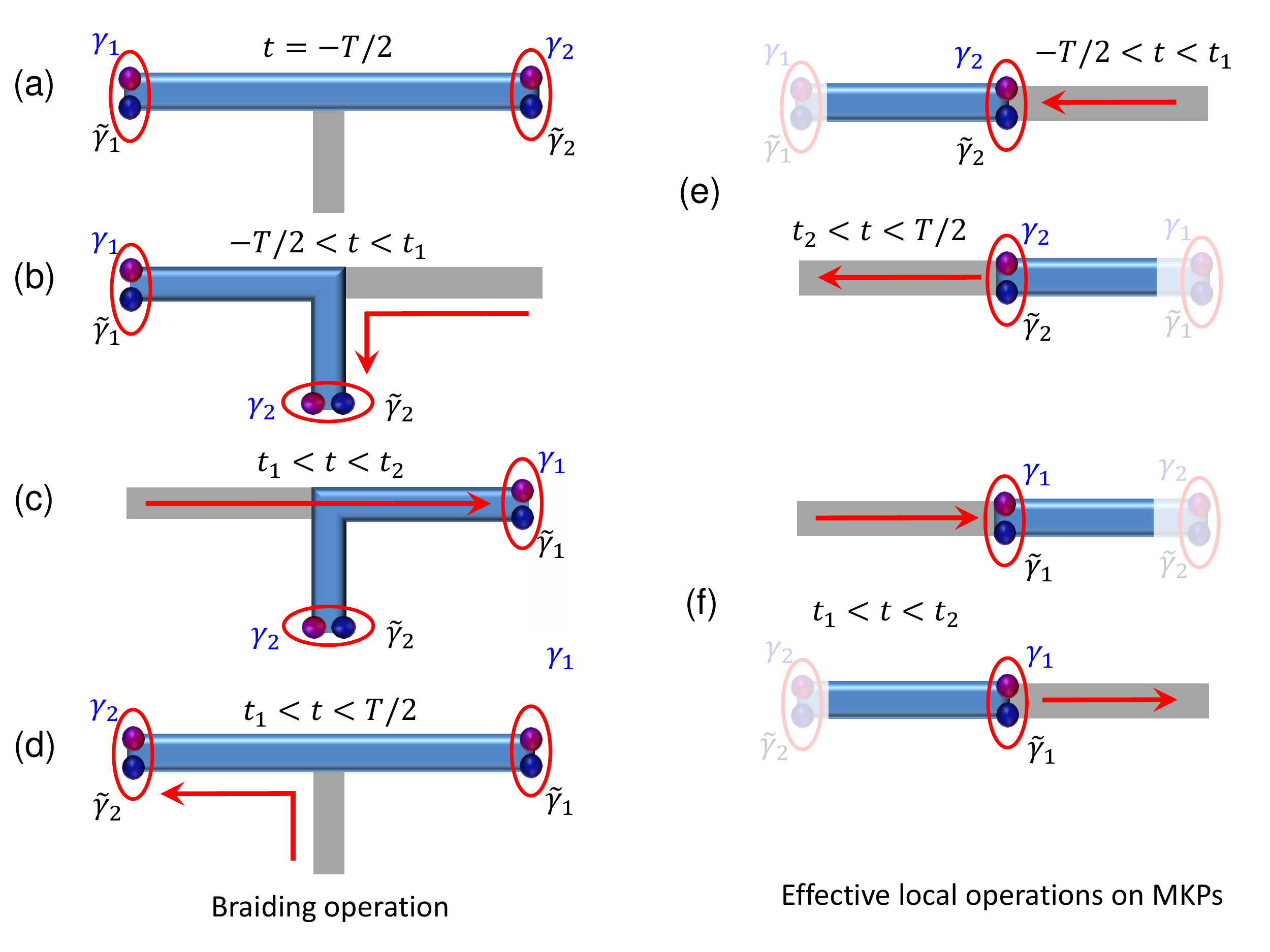} \caption{(Color online) (a-d) Braiding of two MKPs for a single segement of a topological superconductor. The sketched transporting paths in (e) and (f) govern the effective local rotations of the MKPs $\gamma_2,\tilde\gamma_2$ and $\gamma_1,\tilde\gamma_1$, respectively, during the braiding. It can be seen that the two paths are inverse to each other and thus give rise to the opposite local rotations for the two MKPs if the system has no dynamical noise.}\label{swappingSI1}
\end{figure}
\begin{figure}[h]
\includegraphics[width=0.6\columnwidth]{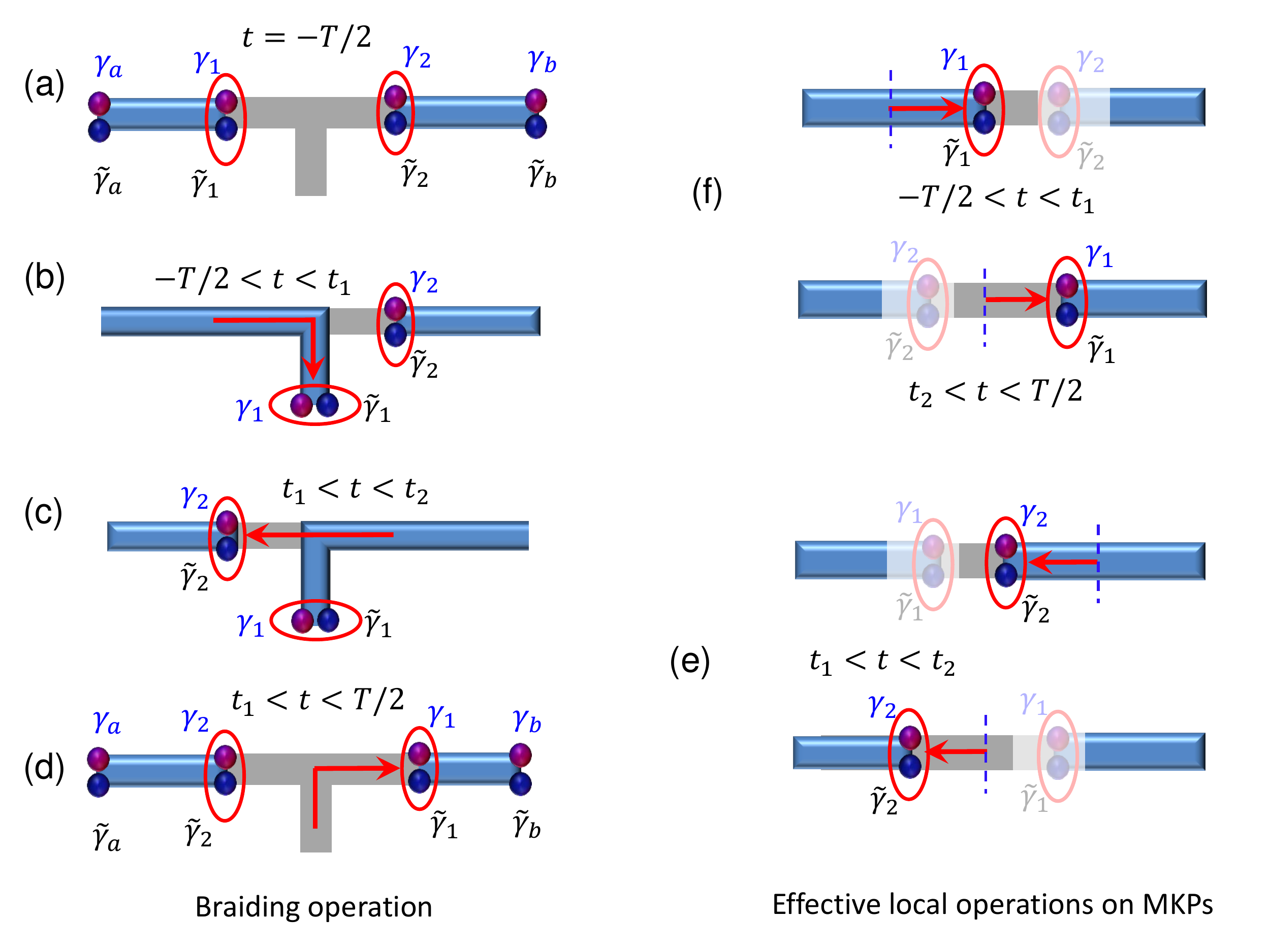} \caption{(Color online) (a-d) Braiding of two MKPs on two separated segments of a topological superconductor. The sketched transporting paths in (e) and (f) govern the effective local rotations of the MKPs $\gamma_2,\tilde\gamma_2$ and $\gamma_1,\tilde\gamma_1$, respectively, during the braiding. Similar to the cases in Fig.~\ref{swappingSI1}, the two paths are inverse to each other and thus give rise to the opposite local rotations for the two MKPs.}\label{swappingSI2}
\end{figure}
In this subsection we show that the Majorana swapping symmetry defined above can always be constructed if the system has no dynamical noise. We first show the generic results, and then present the numerical simulation. Let the total braiding Hamiltonian be denoted as $H(t)=H_0+\delta H(t)$, where $H_0$ is the static part describing the 1D TR invariant TSC, and the time-dependent term $\delta H(t)$ describes the braiding manipulation.
Since the system has no dynamical noise, the time-dependence of $\delta H(t)$ solely comes from the adiabatic parameter manipulation, e.g. the gate control, for the braiding. On the other hand, the static disorder can be generically taken into account in the system. We note that during the braiding manipulation, the static disorder may still induce local rotation on the MKPs. This effect can be understood in the following way. During the braiding manipulation, the MKPs are transported by e.g. gate control. In the co-moving frame, the MKPs experience the variation in time of the local parameters of the Hamiltonian. Such variation may give rise to effective local rotation of the MKPs. As shown below, the key thing is that the local rotations on the MKPs $\gamma_1,\tilde\gamma_1$ and $\gamma_2,\tilde\gamma_2$ are exactly opposite.

Fig.~\ref{swappingSI1} and Fig.~\ref{swappingSI2} show the braiding processes for two MKPs on a single segment and two separated segments of a topological superconductor, respectively. From the sketched manipulations, the local operations on each MKP are described in (e) and (f) of the two figures. The local operation on the vertical wire of the T-junction is cancelled due to the inverse control of the system parameters for a braiding. It can be seen that while the local operations on the MKPs can be generically nonzero, they are opposite for $\gamma_1,\tilde\gamma_1$ and $\gamma_2,\tilde\gamma_2$ by comparing the transporting paths shown in (e) and (f) of Figs.~\ref{swappingSI1} and~\ref{swappingSI2}. In other words, suppose that the local rotation of $\gamma_1,\tilde\gamma_1$ is described by the mixing angle $\theta_b$. The corresponding angle for $\gamma_2,\tilde\gamma_2$ should be $-\theta_b$. As a result, we can construct a Majorana swapping operator $\hat S$, transforming the Hamiltonian by $SH(-t)S^{-1}=H(t)$ and MKPs according to
\begin{eqnarray}
\hat S\gamma_{1/2}\hat S^{-1}=\cos(\pm\theta_b)\gamma_{2/1}+\sin(\pm\theta_b)\tilde\gamma_{2/1}, \ \ \hat S\tilde\gamma_{1/2}\hat S^{-1}=\cos(\pm\theta_b)\tilde\gamma_{2/1}-\sin(\pm\theta_b)\gamma_{2/1}.
\end{eqnarray}
In particular, if the TSC is inversion symmetric with respect to the junction position between the two MKPs, one has $\theta_b=0$. The braiding operator is generically followed by ${\cal B}_{12}=\exp(-\pi\gamma_1\gamma_2'/4)\exp(-\pi\tilde\gamma_1\tilde\gamma_2'/4)$ with $\gamma_2'=\cos\theta_b\gamma_2+\sin\theta_b\tilde\gamma_2$ and $\tilde\gamma_2'={\cal T}\gamma_2'{\cal T}^{-1}$, recovering the non-Abelian braiding.

Now we present the numerical simulation of the above results, which confirms that the non-Abelian braiding is valid for the system without dynamical noise, and is completely immune to the static disorder scatterings. For the numerical simulation, we consider the 1D TR invariant TSC described by
\begin{eqnarray}\label{braidingSI}
H_0&=&\sum_{\left<i,j\right>,\sigma}\! t_0 c_{i\sigma}^\dagger c_{j\sigma}+\sum_{j}\!(\pm\alpha_R c_{j\uparrow}^\dagger c_{j\pm1\downarrow}\!+\Delta_{p} c_{j\uparrow}c _{j+1\uparrow}\nonumber\\
&&+\Delta^*_{p} c_{j\downarrow}c _{j+1\downarrow}+\Delta_s c_{j\uparrow} c_{j\downarrow}\!+\!h.c.)\!-\!\mu\!\sum_{j \sigma} \!n_{j \sigma}
\end{eqnarray}
and random static onsite disorder potential
\begin{eqnarray}\label{disorderSI}
V_{\rm dis}&=&\sum_{j}W_j(n_{j \uparrow}+n_{j\downarrow}),
\end{eqnarray}
with disorder strength $W_j$ randomly distributed within the range of $[-W_0,W_0]$. We note that the other type of disorder, e.g. a random distribution of $s$-wave pairing or hopping coefficients, give the similar results. We take the parameters to be dimensionless and $t_0=10, \Delta_s=1$, $\Delta_p=2$, $\alpha_R=1$. The system can be in topological regime for $|\Delta_p|>|\Delta_s|$. In the topological region the chemical potential is set as $\mu=7$, with the topological gap without disorder being $E_g\approx 0.58$. We demonstrate the braiding through the junction by locally tuning $\mu$ in the configuration shown in Fig.~\ref{swappingSI2}(a-d).
The non-Abelian braiding of MKPs is confirmed by numerically calculating the magnitude $\eta(t)=\langle\gamma_1(0)|\gamma_1(t)\rangle|$, which equals $-1$ after a full braiding at $t=2T$ (similar for other Majorana modes), namely, all the Majorana modes change sign after a full braiding. On the other hand, the adiabatic condition is satisfied in that the projection of instantaneous Majorana state to the initial modes, given by $\zeta(t)=\sum_{j=1,2}[|\langle\gamma_1(t)|\gamma_j(0)\rangle|^2+|\langle\gamma_1(t)|\tilde\gamma_j(0)\rangle|^2]$, returns to unity after a single ($t=T$) and full ($t=2T$) braiding. From the results shown in Fig.~\ref{disorderSI}, we can see that increasing $W_0$ induces more fluctuations in the Majorana wave functions in the intermediate evolution, but does not affect the result after braiding. We have numerically checked many more different parameter conditions, including different configurations of the pairing and chemical potential, and show that the non-Abelian braiding is precisely valid in all the parameter regimes and different configurations of static disorder potentials without destroying the bulk topology of the superconductor. With these results, we have confirmed the non-Abelian braiding of MKPs in the case without dynamical noise.
\begin{figure}[t]
\includegraphics[width=0.6\columnwidth]{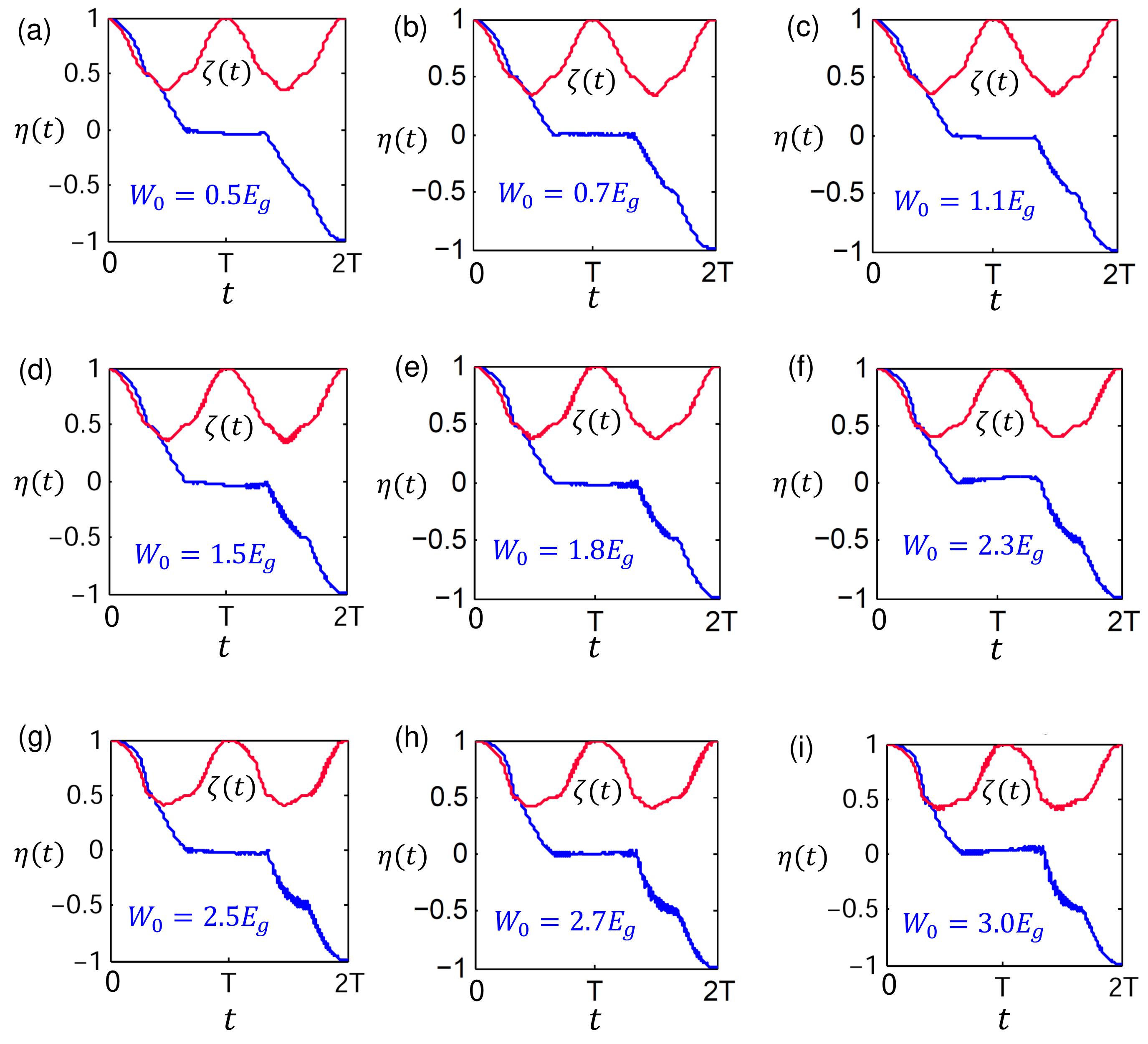} \caption{(Color online) Non-Abelian braiding of MKPs immune to static disorder scattering. The setup sketched in Fig.~\ref{swappingSI2} (1-d) is used for the numerical simulation. (a-i) The evolution of MKP wave function in a full braiding with different disorder strengths $W_0$ ($E_g$ is the bulk gap of the TSC). The non-Abelian braiding of MKPs is confirmed by that the magnitude $\eta(t=2T)=\langle\gamma_1(0)|\gamma_1(t)\rangle|_{t=2T}=-1$ after a full braiding at $t=2T$ (similar for other Majorana modes). The adiabatic condition is satisfied in that the projection $\zeta(t)=\sum_{j=1,2}[|\langle\gamma_1(t)|\gamma_j(0)\rangle|^2+|\langle\gamma_1(t)|\tilde\gamma_j(0)\rangle|^2]$ returns to unity after a single ($t=T$) and full ($t=2T$) braiding.}
\label{disorderSI}
\end{figure}

\section*{II. Dynamical noise}

\subsection*{A. High-order transitions and dynamical time-reversal symmetry}

The dynamical noise may induce random local operations on MKPs if it breaks certain symmetries. Differently from the effect of static disorder scatterings, such local rotations are random and uncorrelated for different MKPs, thus leading to the error of the non-Abelian braiding. For a generic study, we consider a MKP in a 1D TR invariant TSC, denoted as $\gamma_a$ and $\tilde\gamma_a$, coupled to the bulk fermionic modes (denoted as $c_j$ and with energy $\epsilon_{j}$) by a dynamical perturbation. The Hamiltonian can be generically described by $H=H_0+H_{p}$, with
\begin{eqnarray}
H_{0}&=&\sum_{j}2\epsilon_{j}(c_{j}^{\dagger}c_{j}+\tilde{c}_{j}^{\dagger}\tilde{c}_{j})\nonumber\\
H_{p}&=&\gamma_a \sum_{j}2 (V_{j1} c_{j}-V_{j1}^{*} c_{j}^{\dagger} +V_{j2} \tilde{c}_{j}-V_{j2}^{*} \tilde{c}_{j}^{\dagger}) + {\rm T.P.}\nonumber
\end{eqnarray}
where the fermionic modes $\tilde{c}_{j}=\mathcal{T}c_{j}\mathcal{T}^{{-1}}$ and ${\rm T.P.}$ means TR part of the former term in $H_{p}$. The Hamiltonian $H_0$ describes the eigenbases of the bulk superconductor, while $H_{p}$ describes the couplings between bulk fermionic modes and the MKP by the dynamical noise $V_{jm}(t)$. The correlation functions of the dynamical noise are given by
\begin{eqnarray}\label{noiseSI1}
\left\langle V_{j1}(t_1)V_{j2}(t_2) \right\rangle_0&=&V_0^2\mathcal{C}_{j}(t_1-t_2),\\
\hspace{4mm}\sum_{m=1}^2\left\langle V_{jm}(t_1)V_{jm}(t_2) \right\rangle_0&=&V_0^2\mathcal{F}_j(t_1-t_2).
\end{eqnarray}
Here $V_0^2$ characterizes the coupling strength between Majorana and bulk states, $\langle\cdots\rangle_0$ denotes the configuration averaging in time domain, which can be replaced by time averages for an ergodic system
\begin{eqnarray}
\left\langle V_{jm_1}(t_1)V_{jm_2}(t_2) \right\rangle_0=\lim_{T_0\rightarrow\infty}\int_{-T_0}^{T_0}\frac{dt}{2T_0}V_{jm_1}(t)V_{jm_2}(t+t_2-t_1),
\end{eqnarray}
with $m_{1,2}=1,2$, and $\mathcal{C}_{j}(\tau)$ and $\mathcal{F}_{j}(\tau)$ determine the properties of the correlation functions. Besides, we take the configuration averaging to be $\left\langle V_{jm} \right\rangle_0=0$, since any nonzero value of this term can be absorbed by redefining the energy of fermionic states.

The mixing between the Majorana modes of a MKP can be studied by time-dependent perturbation. The time-dependent evolution operator $U(t)={\hat T}\exp(-i\int^t_{-T/2}H(\tau)d\tau)$ reads
\begin{eqnarray}
U(t)&=&\!U_0(t)\bigr[1\!-\! i\int_{-T/2}^t\! d\tau U_0^\dagger(\tau)H_{int}(\tau)U_0(\tau)\!-\nonumber\\
&&-\!\int_{-T/2}^t\!d\tau_1\int_{-T/2}^{\tau_1}\!d\tau_2 U_0^\dagger(\tau_1)H_{int}(\tau_1)U_0(\tau_1)U_0^\dag(\tau_2)H_{int}(\tau_2)U_0(\tau_2)\bigr]\cdots,
\end{eqnarray}
with $U_0(t)=\exp(-i\int^t_{-T/2} H_0d\tau)$.
Expand $H_p$ in the diagonalized Nambu basis $(c_j^\dag, c_j, \tilde c_j^\dag, \tilde c_j,\gamma_a, \tilde\gamma_a)$:
\begin{eqnarray}
H_p&=&\sum_j H_p(j),\\\nonumber
H_p(j)&=&(c_j^\dag, c_j, \tilde c_j^\dag, \tilde c_j,\gamma_a, \tilde\gamma_a)\left(
\begin{array}{cccccc}
0  &0 &0 &0 &v_{j1}^* & v_{j2}\\
0  &0 &0 &0 &-v_{j1} &v_{j2}^*\\
0  &0 &0 &0 &v_{j2}^* &-v_{j1}\\
0  &0 &0 &0 &-v_{j2} &v_{j1}^*\\
v_{j1} &-v_{j1}^* &v_{j2} &-v_{j2}^* &0 &0\\
v_{j1}&^* v_{j1} &-v_{j2}^* &v_{j2} &0 &0\\
\end{array}
\right)(c_j^\dag, c_j, \tilde c_j^\dag, \tilde c_j,\gamma_a, \tilde\gamma_a)^T.\nonumber
\end{eqnarray}
As is shown in $H_p$, the direct coupling between different Majorana modes is missing in $H_p$, thus the 1st order correction to $\chi(t)$ is zero. We first consider the second order correction. The amplitude of transition from one Majorana mode ($\gamma_a$) to another ($\tilde\gamma_a$), calculated by $\chi(t)=\langle\tilde{\gamma}_a |U(t)|\gamma_a\rangle$, is shown to be
\begin{eqnarray}\label{ftranSI}
\chi(t)&=&\int_{T/2}^t\!d\tau_1\int_{T/2}^{\tau_1}\!d\tau_2\left\langle\tilde{\gamma}_0 |U^{(2)}(\tau_1,\tau_2) |\gamma_0\right\rangle_{0}\nonumber\\
&=&\sum_j \int_{T/2}^{t}d \tau_{1}\int_{T/2}^{\tau_1}d \tau_{2}\bigr\langle[-V_{j1}(\tau_1)V_{j2}(\tau_2)+V_{j2}(\tau_1)V_{j1}(\tau_2)]e^{i\epsilon_j(\tau_1-\tau_2)}\nonumber\\
&&+[-V_{j1}^*(\tau_1)V_{j2}^*(\tau_2)\!+\!V_{j2}^*(\tau_1)V_{j1}^*(\tau_2)]e^{-i\epsilon_j(\tau_1-\tau_2)}]\bigr\rangle_{0}\nonumber\\
&=&2V_0^2\sum_j \int_{-T/2}^{t}d \tau_{1}\int_{-T/2}^{\tau_1}d \tau_{2}\ \Re\biggr\{\bigr[\mathcal{C}_j(\tau_1-\tau_2)-\mathcal{C}_j(\tau_2-\tau_1)\bigr]e^{i\epsilon_j(\tau_1-\tau_2)}\biggr\}.
\end{eqnarray}
In the above formula $U^{(2)}(\tau_1,\tau_2)=U_0^\dagger(\tau_1)H_{int}(\tau_1)U_0(\tau_1)U_0^\dag(\tau_2)H_{int}(\tau_2)U_0(\tau_2)$.
The higher order perturbation calculation can be proceeded in a similar way. Note that an odd order perturbation does not contribute to the transition from one Majorana mode to its time-reversal partner, and an even order perturbation can be derived using the results of the second order perturbation calculation.
\begin{eqnarray}
\langle\gamma_a|U^{(2)}(\tau_1,\tau_2)|\gamma_a\rangle&=&\sum_j\Re\biggr\{\bigr[V_{j1}^*(\tau_1)V_{j1}  (\tau_2)+V_{j2}^*(\tau_1)V_{j2}(\tau_2)]e^{i\epsilon_j(\tau_1-\tau_2)}\biggr\}=\sum_j\Re\biggr\{ A_j(\tau_1,\tau_2)\biggr\},\label{virtuallySI1}\\
\langle\gamma_a|U^{(2)}(\tau_1,\tau_2)|\tilde\gamma_a\rangle&=&\sum_j\Re\biggr\{\bigr[V_{j1}^*(\tau_1)V_{j2}^*(\tau_2)-V_{j2}^*(\tau_1)V_{j1}^*(\tau_2)]e^{i\epsilon_j(\tau_1-\tau_2)}\biggr\}=\sum_j\Re\biggr\{ B_j(\tau_1,\tau_2)\biggr\},\label{virtuallySI2}\\
\langle\tilde\gamma_a|U^{(2)}(\tau_1,\tau_2)|\gamma_a\rangle&=&\sum_j\Re\biggr\{\bigr[V_{j2}(\tau_1)V_{j1}(\tau_2)-V_{j1}(\tau_1)V_{j2}(\tau_2)]e^{i\epsilon_j(\tau_1-\tau_2)}\biggr\}=\sum_j\Re\biggr\{ C_j(\tau_1,\tau_2)\biggr\},\label{virtuallySI3}\\
\langle\tilde\gamma_a|U^{(2)}(\tau_1,\tau_2)|\tilde\gamma_a\rangle&=&\sum_j\Re\biggr\{[V_{j1}(\tau_1)V_{j1}^*(\tau_2)+V_{j2}(\tau_1)V_{j2}^*(\tau_2)]e^{i\epsilon_j(\tau_1-\tau_2)}\biggr\}=\sum_j\Re\biggr\{ D_j(\tau_1,\tau_2)\biggr\}.\label{virtuallySI4}
\end{eqnarray}
Eqs.~\eqref{virtuallySI1} and~\eqref{virtuallySI4}  describe a process when a Majorana mode ($\gamma_a$/$\tilde\gamma_a$) transitions virtually to a bulk fermionic state (e.g. $c_j$), and then transitions back to its former state ($\gamma_a$/$\tilde\gamma_a$), while Eqs.~\eqref{virtuallySI2} and~\eqref{virtuallySI3} describe a process when a Majorana mode ($\gamma_a$/$\tilde\gamma_a$) transitions virtually to a bulk fermionic state (e.g. $c_j$), and then transitions back to its TR partner state ($\tilde\gamma_a$/$\gamma_a$).
Inserting identity matrice into the higher order perturbation calculation can efficiently simplify the calculation. We have
\begin{eqnarray}
\langle\tilde\gamma_a|U^{(2m)}(\tau_1,\tau_2)|\gamma_a\rangle
&=&\int_{-T/2}^t\!d\tau_1\int_{-T/2}^{\tau_1}\!d\tau_2\cdots\int_{-T/2}^{\tau_{2m-2}}\!d\tau_{2m-1}\int_{-T/2}^{\tau_{2m-1}}\!d\tau_{2m} \nonumber\\
&\times& \langle\tilde\gamma_a|U_0^\dagger(\tau_1)H_{int}(\tau_1)U_0(\tau_1)U_0^\dag(\tau_2)H_{int}(\tau_2)U_0(\tau_2)\sum_{n}|n\rangle\nonumber\\
&\times&\langle n| U_0^\dagger(\tau_1)H_{int}(\tau_1)U_0(\tau_1)U_0^\dag(\tau_2)H_{int}(\tau_2)U_0(\tau_2)\sum_{n}|n^{\prime}\rangle\times\cdots\nonumber\\
&\times&\langle n^{\prime\prime}|U_0^\dagger(\tau_1)H_{int}(\tau_1)U_0(\tau_1)U_0^\dag(\tau_2)H_{int}(\tau_2)U_0(\tau_2)|\gamma_a\rangle,\label{highorderSI}
\end{eqnarray}
where the identity matrix $I=\sum_n|n\rangle\langle n|$, with $|n\rangle$s being the eigenstates of the unperturbed BdG Hamiltonian $H_0$. A second order transition survives only when the initial and final states are both bulk states and when they are both Majorana states. Thus the summation over all the eigenstates of $H_0$ including all the bulk states and Majorana states reduces into the summation over all the Majorana states and Eq.~\eqref{highorderSI} becomes:
\begin{eqnarray}
&&\langle\tilde\gamma_a|U^{(2m)}(\tau_1,\tau_2)|\gamma_a\rangle\nonumber\\
&=&\sum_{j_1}\cdots \sum_{j_m}\int_{-T/2}^t\!d\tau_1\int_{-T/2}^{\tau_1}\!d\tau_2\cdots\int_{-T/2}^{\tau_{2m-2}}\!d\tau_{2m-1}\int_{-T/2}^{\tau_{2m-1}}\!d\tau_{2m} \nonumber\\
&& \times{\sum}^\prime\Re\biggr\{\mathcal{A}_1(\tau_1,\tau_2)\biggr\}\cdots\Re\biggr\{\mathcal{A}_m(\tau_{2m-1},\tau_{2m})\biggr\},
\end{eqnarray}
with $\mathcal{A}_{1\cdots m}=A,B,C,D$ and ${\sum}^\prime$ denoting the summation over all the possible $2m$-th order transitions. To accomplish a $2m$-th order transition with $\gamma_a$ being its initial state and $\tilde\gamma_a$ its final state, Term C always appears one more time than Term B. On the contrary, for a $2m$-th order transition with with $\tilde\gamma_a$ being its initial state and $\gamma_a$ its final, Term B always appears one more time than Term C. While for an arbitrary $2m$-th order transition with the same initial and final states, Term B appears just as often as Term C. In the first case, which is what we are concerned about here, Term C, which describes a second order transition from one Majorana mode ($\gamma_a$) to its time-reversed partner ($\tilde\gamma_a$), appears at least once.
From the above formula we can find two important features of the dynamical noise. Firstly, if the noise correlation respects a {\it dynamical TR symmetry} in the time domain, namely,
\begin{eqnarray}
\langle V_{j_1\nu}(\tau_1)V_{j_1\nu}(\tau_2)\cdots V_{j_m1}(\tau_{2m-1})V_{j_{m2}}(\tau_{2m})\cdots \rangle_0=\langle V_{j_1\nu}(\tau_1)V_{j_1\nu}(\tau_2)\cdots V_{j_m1}(\tau_{2m})V_{j_{m2}}(\tau_{2m-1})\cdots \rangle_0,
\end{eqnarray}
the coupling between $\gamma_a$ and $\tilde\gamma_a$ vanishes, and the Majorana swapping symmetry is
guaranteed. {\it This is a profound result which generalizes the TR symmetry protection of MKP to the dynamical regime.} Specially, the dynamical symmetry can be simply written down as
\begin{eqnarray}
\mathcal{C}_j(\tau)=\mathcal{C}_j(-\tau).
\end{eqnarray}
Accordingly, to have nonzero couplings within a MKP, the correlation function must break the above dynamical TR symmetry.
Secondly, the leading order transition probability in a MKP is proportional to $V_0^4$, namely
\begin{eqnarray}
{\cal D}=|\chi(T)|^2\propto V_0^4/E_{g}^4+{\cal O}(V_0^8/E_{g}^8).
\end{eqnarray}
This implies that the leading order contribution due to dynamical noise is a second-order transition, which describes the process that a Majorana mode ($\gamma_a$) transitions virtually to a bulk fermionic state (e.g. $c_j$), and then transitions back to its TR partner ($\tilde\gamma_a$). This result is deeply connected to the TR symmetry of the original DIII class TSC, which excludes the first-order direct coupling in a MKP. Actually, one can easily show that ${\cal T}\chi(t){\cal T}^{-1}=-\chi(t)$, which implies that the second-order indirect transition effectively breaks the TR symmetry if the correlation function ${\cal C}_j(\tau)$ is not symmetric in time domain. Note that the dynamical noise can also bring about decoherence effect in the Majorana qubits in a D-class chiral topological superconductor without TR symmetry, where the leading contribution comes from the first-order transition~\cite{Goldstein2011}. The high-order contribution shows that while the dynamical noise may be detrimental, the caused error can be generically reduced when the noise strength is not strong.

\subsection*{B. Suppression of local mixing for random dynamical noise}

For a quantitative study, we consider the Hamiltonian for the TSC and dynamical noise in the forms
\begin{eqnarray}
H_0&=&\sum_{\left<i,j\right>,\sigma}\! t_0 c_{i\sigma}^\dagger c_{j\sigma}+\sum_{j}\!(\pm\alpha_R c_{j\uparrow}^\dagger c_{j\pm1\downarrow}\!+\Delta_{p} c_{j\uparrow}c _{j+1\uparrow}\nonumber\\
&&+\Delta^*_{p} c_{j\downarrow}c _{j+1\downarrow}+\Delta_s c_{j\uparrow} c_{j\downarrow}\!+\!h.c.)\!-\!\mu\!\sum_{j \sigma} \!n_{j \sigma},\\
%\begin{split}
H_{p}&=&\sum_{j}V_j[\cos(\omega t)(c_{j\uparrow} c_{j\downarrow}\!+\!h.c.)\!-\!\cos (\omega t+\delta \phi_j)n_j].
%\end{split}
\end{eqnarray}
The noise Hamiltonian $H_p$ describes the dynamical fluctuations of frequency $\omega$ in the $s$-wave pairing and chemical potential (the fluctuations in other parameters, e.g. the SO coupling and $p$-wave pairing, are similar), with $n_j=n_{j\uparrow}+n_{j\downarrow}$. The amplitude and phase difference $\delta \phi_j$ in fluctuating chemical potential and $s$-wave pairing can in general be position dependent.

Without loss of the generality, we write down the wave functions of the four Majorana modes $\gamma_1,\tilde\gamma_1,\gamma_2$, and $\tilde\gamma_2$ by
\begin{eqnarray}
\gamma_1&=&\sum_j(u_{1j}c_{j\uparrow}+v_{1j}c_{j\downarrow})+\sum_j(u^*_{1j}c^\dag_{j\uparrow}+v^*_{1j}c^\dag_{j\downarrow}),\\
\tilde\gamma_1&=&\sum_j(u^*_{1j}c_{j\downarrow}-v^*_{1j} c_{j\uparrow})+\sum_j(u_{1j}c^\dag_{j\downarrow}-v_{1j}c^\dag_{j\uparrow}),\\
\gamma_2&=&\sum_j(u_{2j}c_{j\uparrow}+v_{2j}c_{j\downarrow})+\sum_j(u^*_{2j}c^\dag_{j\uparrow}+v^*_{2j}c^\dag_{j\downarrow}),\\
\tilde\gamma_2&=&\sum_j(u^*_{2j}c_{j\downarrow}-v^*_{2j} c_{j\uparrow})+\sum_j(u_{2j}c^\dag_{j\downarrow}-v_{2j}c^\dag_{j\uparrow}),
\end{eqnarray}
where $u_{1j},v_{1j},u_{2j}$, and $v_{2j}$ are superposition coefficients. Rewriting the noise Hamiltonian in terms of the above Majorana modes yields
\begin{eqnarray}\label{Hp}
H_{p}&\approx&\frac{1}{2}\sum_{j}\biggr\{V_j\cos(\omega t)\bigr[\gamma_1(u_{1j}^*c_{j\downarrow}-v_{1j}^* c_{j\uparrow})-
\tilde\gamma_1(v_{1j}c_{j\downarrow}-u_{1j} c_{j\uparrow})\nonumber\\
&&+\gamma_2(u_{2j}^*c_{j\downarrow}-v_{2j}^* c_{j\uparrow})-
\tilde\gamma_2(v_{1j}c_{j\downarrow}-u_{1j} c_{j\uparrow})\bigr]+h.c.\biggr\}\nonumber\\
&&-\frac{1}{2}\sum_j\biggr\{V_j\cos (\omega t+\delta \phi_j)\bigr[\gamma_1(u_{1j}c_{j\uparrow}+v_{1j}c_{j\downarrow})-
\tilde\gamma_1(v_{1j}^*c_{j\uparrow}-u_{1j}^*c_{j\downarrow})\nonumber\\
&&+\gamma_2(u_{2j}c_{j\uparrow}+v_{2j}c_{j\downarrow})-
\tilde\gamma_2(v_{2j}^*c_{j\uparrow}-u_{2j}^*c_{j\downarrow})
\bigr]+h.c.\biggr\}.
\end{eqnarray}
Now we can examine the local coupling between $\gamma_1$ and $\tilde\gamma_1$ induced by the dynamical noise through the second order transition (the coupling between $\gamma_2$ and $\tilde\gamma_2$ is similar). The noise potential $V_j\cos(\omega t+\delta\phi_j)=V_j\cos\omega t\cos\delta\phi_j+V_j\sin\omega t\sin\delta\phi_j$. As shown previously, the local couplings require that the noise correlation function breaks the dynamical TR symmetry, for which only the second-order terms proportional to $\langle \cos(\omega t_1)\sin(\omega t_2)\sin\delta\phi_j\rangle_0\bigr[|u_{1j}|^2\langle c_{j\downarrow}(t_1)c_{j\downarrow}^\dag(t_2)\rangle+|v_{1j}|^2\langle c_{j\uparrow}(t_1)c_{j\uparrow}^\dag(t_2)\rangle\bigr]$ contribute to the local mixing in a MKP. Accordingly, if $\delta\phi_j=0$ over the system, which gives $H_p(t)=H_p(-t)$, the coupling in a MKP vanishes and the local rotations disappear. With these results in mind and from the above formula~\eqref{Hp} we can show that
\begin{eqnarray}
{\cal D}(T)\approx1-\cos^2\bigr[\frac{\langle\gamma_a|V_j|\gamma_a\rangle^2\omega T}{12(E_g^2+\langle\gamma_a|V_j|\gamma_a\rangle^2)}\langle\gamma_a|\sin\delta\phi_j|\gamma_a\rangle\bigr],
\end{eqnarray}
with $a=1,2$ and
\begin{eqnarray}
\langle\gamma_a|V_j|\gamma_a\rangle=\sum_jV_j(|u_{1j}|^2+|v_{1j}|^2), \ \
\langle\gamma_a|\sin\delta\phi_j|\gamma_a\rangle=\sum_j\sin\delta\phi_j(|u_{1j}|^2+|v_{1j}|^2)
\end{eqnarray}
being the weighted averaging of the noise potential and phase within the Majorana localization length.
The noise frequency is assumed to be $\omega=3/T\ll E_g$, and for weak noise regime the results can be approximated by
\begin{eqnarray}
\mathcal{D}\approx\frac{\langle\gamma_a|V_j|\gamma_a\rangle^4}{16E_g^4}\langle\gamma_a|\sin\delta\phi_j|\gamma_a\rangle^2
-\frac{\langle\gamma_a|V_j|\gamma_a\rangle^8}{763E_g^8}\langle\gamma_a|\sin\delta\phi_j|\gamma_a\rangle^4.
\end{eqnarray}
The above formula shows that the error induced by dynamical noise is small if the noise strength is not strong. Moreover, the effect of dynamical noise can be further greatly suppressed when the noise phase or amplitude has a random distribution within the MKP localization length $\lambda_M$. In this case, the symmetry conditions of the non-Abelian braiding are approximately recovered by average, and the non-Abelian braiding of MKPs is validated. We note that the dynamical noise can also bring about decoherence for Majorana qubit states in the TR symmetry breaking SCs, where the leading order contribution comes from the first order transition. In comparison, the high-order contribution for the TR invariant TSCs implies that while the dynamical noise may be detrimental, the caused error is negligible for the weak noise regime.

%%%%%%%%%%%%%%%%%%%%%%%%%%%%%%%%%%%%%%%%%%%%%%%%%%%%%%%%%%%%%%%%%%%%%%%%%%%%%

 \bibliographystyle{apsrev}
%\bibliography{Liu_bibliography,C:/Users/Alejandro/alobos/mybib/totphys,C:/Users/Alejandro/alobos/mybib/mybibliography}

%\bibliography{D:/mybib/Liu_bibliography,D:/mybib/totphys,D:/mybib/mybibliography}

\begin{thebibliography}{44}
\bibitem{note} In a real Majorana quantum wire, the typical localization length of a Majorana mode can be hundreds of $nm$ to several $\mu m$.
\end{thebibliography}

\end{document}